# Global pattern and mechanism of terrestrial evapotranspiration change indicated by weather stations

First author: Haiyang Shi

**Abstract**

Accurate estimation of global terrestrial evapotranspiration (ET) is essential to understanding changes in the water cycle, which are expected to intensify in the context of climate change. Current global ET products are derived from physics-based, yet empirical, models, water balance methods, or upscaling from sparse in situ observations. However, these products contain substantial limitations such as the coarse resolution due to the coarse climate reanalysis forcing data, the assumptions on the parameterization of the process, the sparsity of the observations, and the lack of global accuracy validation. Using estimates of ET based on the global weather station network and machine learning, we show that global ET ranged from 493 to 522 mm yr$^{-1}$ and increased at the rate of 0.60 mm yr$^{-2}$ from 2003 to 2019. Between the two periods of 2003-2010 and 2011-2019, 61.7% of stations showed an increase in ET. At the large river basin scale, the reliability of the produced ET in this study is comparable to gridded ET data and even higher in regions where weather stations are relatively dense and more representative. Correlation analysis and causal network analysis showed that the main drivers of ET long-term changes are changes in air temperature, radiation, vegetation conditions, and vapor pressure deficit. There is great variability in the causal mechanisms of ET change across vegetation cover and across seasons. This study highlights the promise of using weather stations to complement global ET and water cycle studies at the station scale.

**Main**

Terrestrial evapotranspiration (ET) is one of the most important indicators of the land surface water and energy balance in the global water cycle. ET change is expected to intensify[1,2] in the context of climate change. To better understand the pattern and mechanism of ET change, accurate ET estimation is crucial. Various ET products[3–6] have been widely used in ET-associated water resources[7–9], ecosystem functioning[10,11], carbon and climate feedback[12,13], and agricultural management[14] studies. However, at present, there exists a large uncertainty in global ET products[15], which has led to considerable variation in the global ET patterns (Supplementary Table 1) and an inconsistent understanding of the mechanisms of ET change.

Currently, global-scale ET data are mainly derived from empirically parameterized physics-based models (e.g., Penman-Monteith-based[16,17], Priestley-Taylor-based[18]), water balance methods using precipitation (P) minus terrestrial water storage (TWS) and runoff (Q)[19], and upscaling from flux station observations by machine learning[4,5]. Their shortcomings include (i) the empirical parameterization of the water supply regulation (surface resistance or other related surface components) and some of their assumed stationarity in time (e.g., no changes in soil moisture impact over time), (ii) a lack of validation in global applications and extrapolation for each global grid, and (iii) the low spatial resolution (Supplementary Table 1)[3,5,16–21]. More

specifically:

i. The difficulty in estimating surface and aerodynamic resistances has caused considerable uncertainties in ET process modeling[21]. Physics-based ET models must trade off the physical realism and determination of key parameters such as canopy resistance[16,17,21], and especially its dependence on soil moisture. Recently, emerging hybrid physics-machine learning techniques have been used to represent these complex processes in a data-driven manner without complex parameterizations and have been proven more effective[4], but have not yet been used globally.

ii. The calibration and verification of either physical or statistical ET models are predominantly based on limited flux station observations (Supplementary Table 1). However, those ET models are not necessarily fully applicable when extrapolated to the global scale, and further evaluation is still needed. This is because there is large temporal and spatial heterogeneity in the ET mechanism across different biomes. To date, an accurate ET dataset at the global scale with good extrapolation capacity is still challenging to obtain and therefore desired.

iii. The uncertainty induced by the coarse-resolution inputs is considerable: ET is typically dependent on spatial heterogeneity and local climates, especially on complex land surfaces[22]. However, in physics-based[16–18] and data-driven[4,5] ET model applications, the ~10 km resolution of climate reanalysis input can be inconsistent with the small scale of local climate and water-heat conditions at flux stations. The validity of the empirical site-scale water-heat flux equations can therefore be reduced. ET obtained by inputting a single value from a spatially averaged 10 km grid point (containing huge spatial heterogeneity) can be very different from ET obtained by first inputting high-resolution or small-scale data into a model and then spatially averaging it. Consequently, the output ET data reliability will decline and studies on the driving mechanisms of global ET trends and patterns based on these products can lead to substantial inconsistency and uncertainty[5,6,23–25].

Given these above-mentioned major limitations in current ET products, this study uses machine learning to produce more accurate ET estimates using weather station data and its sub-diurnal information, which reflects changes in the boundary layer tightly connected to land-surface heat flux partitioning[26]. The fitted model is then used to provide estimates of ET across a global network of weather stations. There are several advantages of this approach. First, meteorological observations of weather stations are at the same scale as those of flux stations, thus avoiding coarse-scale output and achieving high accuracy in a data-driven machine learning framework[27]. Second, based on the promise of ET estimates based on weather stations and boundary layer theory[28–30], this study further incorporates surface soil moisture (SSM) and leaf area index (LAI) to improve the representation of the land surface condition. Third, the spatial and temporal coverage of ET of weather stations can be sufficient (at least much more than flux stations) to achieve a more global assessment and understanding of ET patterns and their driving mechanisms.

Specifically, here we used machine learning to estimate ET using weather station data input to replicate collocated ET from 181 flux towers and then used this to estimate ET across a global network of weather stations (Fig. 1). Further, to reach a quantitative understanding of the projected prediction accuracy in the subsequent global weather station extrapolation and assess the related predictability and applicability of prediction model, we constructed a classifier (Fig. 1b) for identifying stations with low and high accuracy relying on the relations between the accuracy metric correlation coefficient (R) of observed ET and predicted ET and the similarity to training data (assuming that stations that are not similar to the training set typically correspond to low $R^{31,32}$). Stations with R > 0.7 were regarded as high-accuracy cases and stations with R <= 0.7 were regarded as low-accuracy cases. From the leave-one-site-out cross-validation[33] of the flux stations (see Methods), we collected 181 such paired 'R-similarity' records to construct this R classifier. The R classifier was then used to project if R > 0.7 for each weather station. We then screened weather stations with a projected R > 0.7 for the subsequent ET pattern and trend analysis. Finally, based on the screened ET data, we investigated (i) the global terrestrial ET patterns indicated by global weather stations, and (ii) the global sensitivity of ET changes to environmental drivers via the Bayesian Network (BN) which found effective in diagnosing causal relationships in the ecosystem function variation attribution[34] (see Methods). The potential accuracy and global coverage of ET at weather stations are likely to be intermediate between the precision and global coverage of flux network observations and estimates based on reanalyzed data (Fig. 1c). Therefore, this study may update some understanding of global ET changes from the perspective of weather stations.

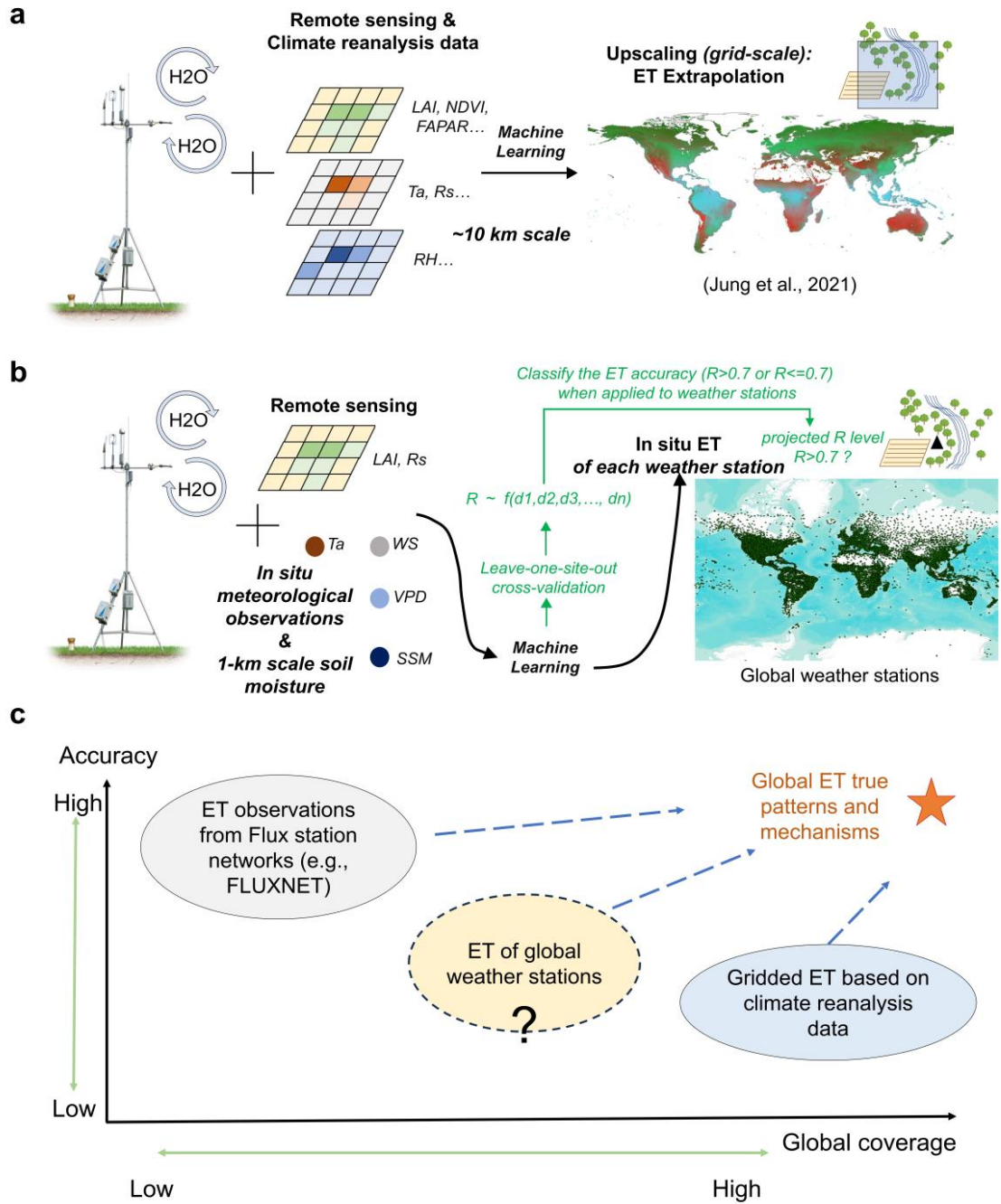

**Fig. 1.** The framework for machine-learning-based evapotranspiration (ET) modeling using the global flux tower observations. **a**, using climate reanalysis as input and upscaling ET to ~10 km resolution grids of the globe[4]. **b**, the ET prediction framework at weather stations proposed in this study aimed to considerably increase the accurate site-scale ET data availability to support future research requiring higher accuracy and higher spatial and temporal coverage. The projected accuracy of the ET time-series data for each weather station is from the classifier (i.e., R ~ f($d_1$, $d_2$, $d_3$, …, $d_n$)) of the R records (obtained in the leave-one-site-out cross-validation) and the distance (i.e., $d_1$, $d_2$, $d_3$, …, and $d_n$) of the predictors of the validation set to the training set (see Methods). **c**, possible variations in accuracy and global coverage of ET based on flux network observations, weather station estimates, and grid reanalysis data-based estimates.

## Weather stations' ET pattern and representativeness

Over 2003 to 2019, global ET (ET_WS) which weighted from ET estimates of weather stations by land cover type area (see Methods) increased by about 0.60 mm yr$^{-2}$ (*p*-value < 0.1) with annual mean ET ranging from 1.35 to 1.43 mm/d (i.e., 493 to 522 mm/yr) (Fig. 2). Such ET value and increasing rate both fall within the range of variation of the previous estimates (Supplementary Table 1). This confirmed the slow increase of terrestrial ET in the context of a warming climate. The highest ET was observed in 2018. From 2003-2010 to 2011-2019, 61.7% of weather stations showed an increase in ET (Fig. 2). In stations showed significant ($p < 0.1$) change in ET, 63.8% showed an increase (Fig. 4). In the cross-product comparison with other estimates of global terrestrial ET, the dynamic trend of ET_WS is partially consistent with that of MOD16A2 and FLUXCOM but with higher interannual variability (Supplementary Fig. 1).

In terms of the ability to close the water budget, the correlation coefficient between ET_WS and ET_WB (calculated from the basin water balance P - Q - ΔTWS, see Methods) reaches 0.671 ($p < 0.001$) in 189 global large river basins (Fig. 3). It is comparable to ET_MOD16's performance of 0.676 ($p < 0.001$) but slightly lower than ET_FLUXCOM's 0.723 ($p < 0.001$). Given the considerable variation of weather station density (Fig. 3a), we also compared the ET product difference in the basin group with relatively high station density. In 53 basins that simultaneously meet relatively high site density and land cover representativeness (see Methods), correlation of ET_WS even showed a higher correlation with ET_WB than ET_MOD16 and ET_FLUXOM (Fig. 3c). This suggests that even after weighting by land cover type, a treatment that can introduce large errors, the resulting ET_WS is still as representative as ET_MOD16 and FLUXCOM for global ET. This confirmed the accuracy of the produced ET data at weather stations. Where station density and land cover type representations are high, ET_WS may have achieved higher accuracy than existing ET products and has the potential to be regarded as a powerful basis for analyzing the mechanisms driving ET changes.

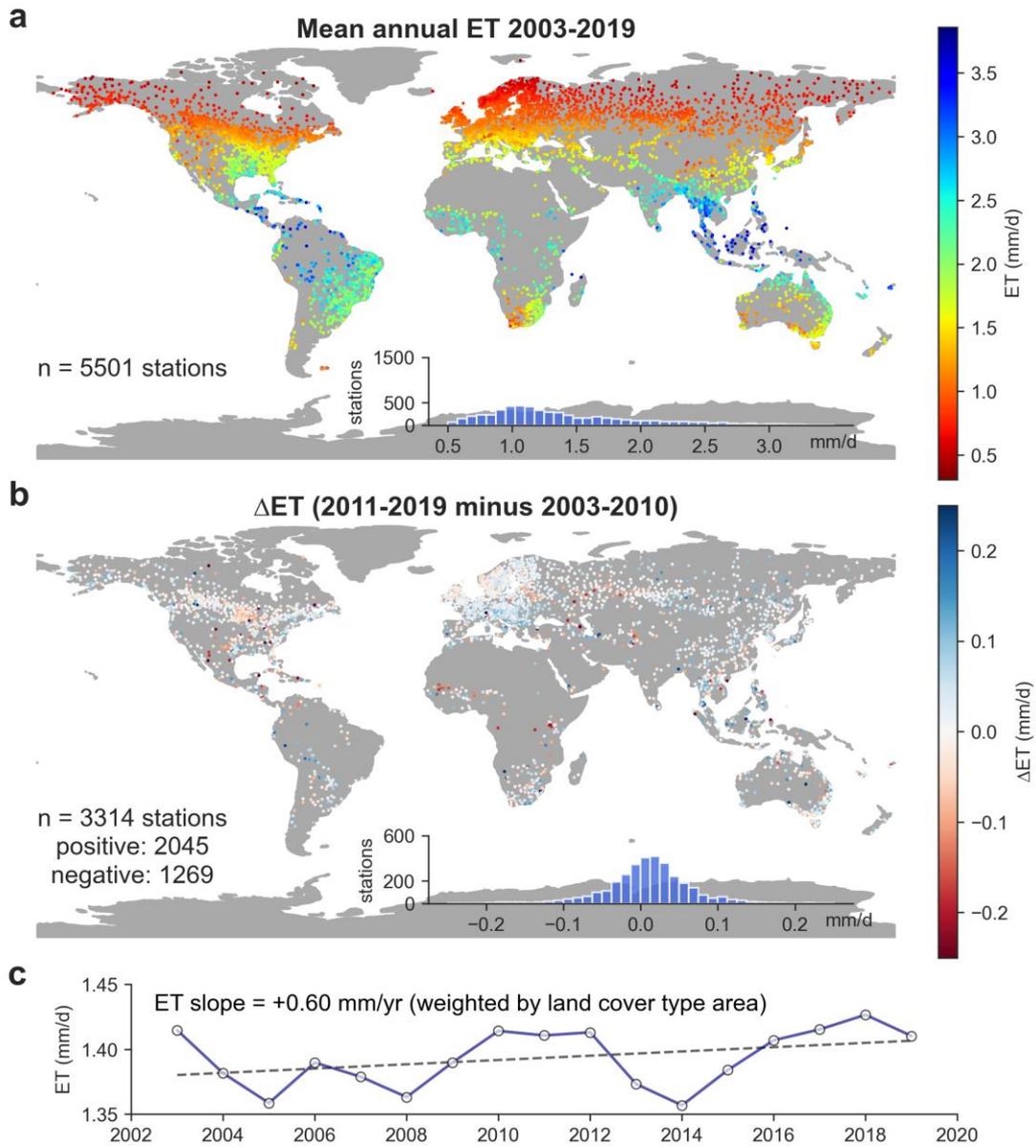

**Fig. 2.** Global ET patterns and trends at weather stations. **a**, estimated mean daily ET at global weather stations from 2003 to 2019. **b**, ΔET as the change in ET from the 2011-2019 period and the 2003-2010 period. **c**, the trend in global ET and its slope weighted by the land cover type area (see Methods).

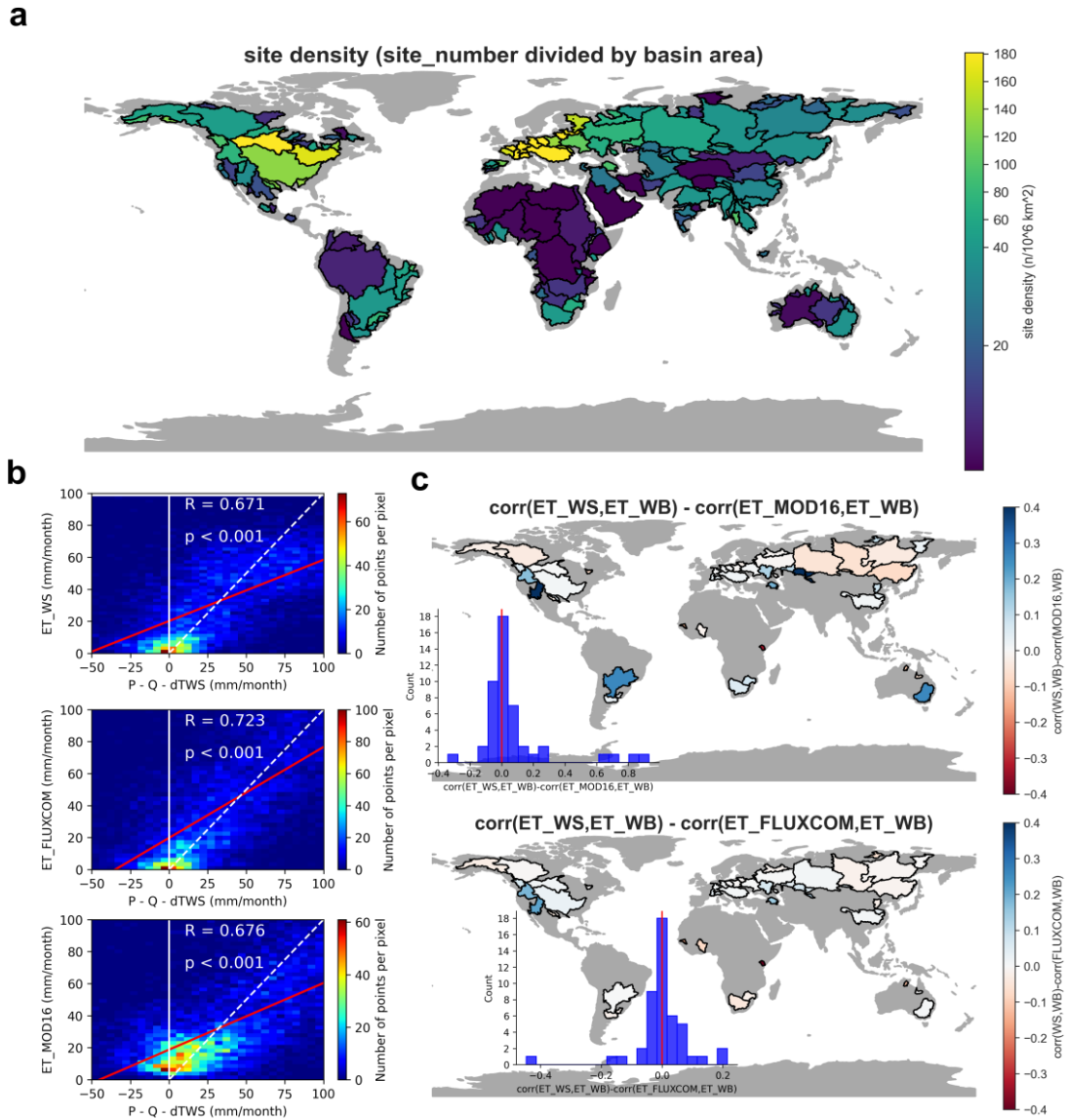

**Fig. 3.** Performance evaluation of the ET of weather stations in representing regional ET. **a**, density of weather stations with accurate (R > 0.7) ET predictions in global 189 large river basins. **b**, correlation of P-Q-ΔTWS (equals ET_WB) with ET_WS, ET_FLUXCOM, and ET_MOD16 respectively. **c**, the difference in correlation of ET_WB with ET_WS compared to ET_FLUXCOM and ET_MOD16 in river basins that simultaneously meet relatively high site density and land cover representativeness (see Methods).

## Casual driving mechanism of ET change

Current findings of studies on the attribution of changes in global ET are not fully consistent and remain uncertain[5,6,23–25], with many factors (e.g. climate change, vegetation conditions change, etc.) thought to have a complex direct and indirect role on ET changes. Here we briefly analyzed the patterns of ET change (ΔET) over two separate periods, from 2003-2010 and 2011-2019, and their possible relations to changes in environmental driver variables (Fig. 5 and Supplementary Fig. 2). At global weather stations, LAI has increased consistently and substantially, which agrees with reported global greening[35]. Daily mean temperature (Ta) is increasing across most of the globe with global warming (except in regions with strong land management or where decadal climate variability in temperature is negative), downward shortwave radiation (RSDN) increased slightly, vapor pressure deficit (VPD) also increased. Spatially, these trends also have strong regional variations (Fig. 4a).

ΔET showed strong significant positive correlations (Fig. 4b) with ΔRSDN (R = 0.58, p < 0.001) and ΔTa (R = 0.57, p < 0.001), and relatively strong positive correlations with ΔLAI (R = 0.25, p < 0.001) and ΔVPD (R = 0.23, p < 0.001). It also exhibited a lower but significant positive correlation with ΔWS (R = 0.10, p < 0.001). ΔET shows a lower but significant negative correlation with ΔSSM (R = -0,08, p < 0.001) and ΔP (R = -0.07, p < 0.001). This suggests that elevated RSDN and Ta dominated the ET increase from 2003 to 2019. Elevated VPD and LAI also contributed to the increase in ET, with VPD possibly increasing ET by increasing atmospheric moisture demand and elevated LAI possibly altering surface water partitioning thus increasing ET. In addition, compared to RSDN, Ta, LAI, and VPD, SSM and P are more likely to fluctuate on an annual scale instead of substantially change at the decadal scale[38], thus the influence of SSM and P can be higher on interannual variability than the decadal variability.

In terms of seasonal characteristics (Supplementary Fig. 3, Supplementary Fig. 4, and Supplementary Fig. 5), compared with June, July and August (JJA), ΔLAI and ΔT showed stronger correlations with ΔET in March, April and May (MAM) and September, October and November (SON). It can be caused by the warming in spring and fall and the extending of the phenological period in early spring and later autumn. Compared to JJA, ΔVPD and ΔSSM also showed slightly stronger correlations with ΔET in MAM and SON. It suggests that the role of water demand increase leading to ET increase is greater at MAM and SON.

Compared to lower vegetation cover (monthly maximum LAI <= 2), stations with higher vegetation cover (monthly maximum LAI > 2), ΔLAI in SON showed a stronger correlation with ΔET (Supplementary Fig. 6). It implies that the positive effects of the extension of the phenological period on ET in SON occurred more at high-vegetation stations. At stations with higher vegetation cover, ΔLAI in JJA showed a weaker correlation with ΔET. Given that the ΔLAI of JJA is less related to phenological changes and more correlated with the increase in $CO_2$ fertilization, the positive response of ET to $CO_2$ fertilization at high-vegetation sites is weaker than at low-vegetation sites. In the context of elevated $CO_2$ concentration, greening (increased vegetation cover and root growth) at low-vegetation stations more frequently led to ET increase[15]. In contrast, in the high-vegetation stations where physiological responses dominated, the decrease in stomatal conductance due to elevated $CO_2$, which can decrease ET, may have partially offset the positive effect of greening on ET[15].

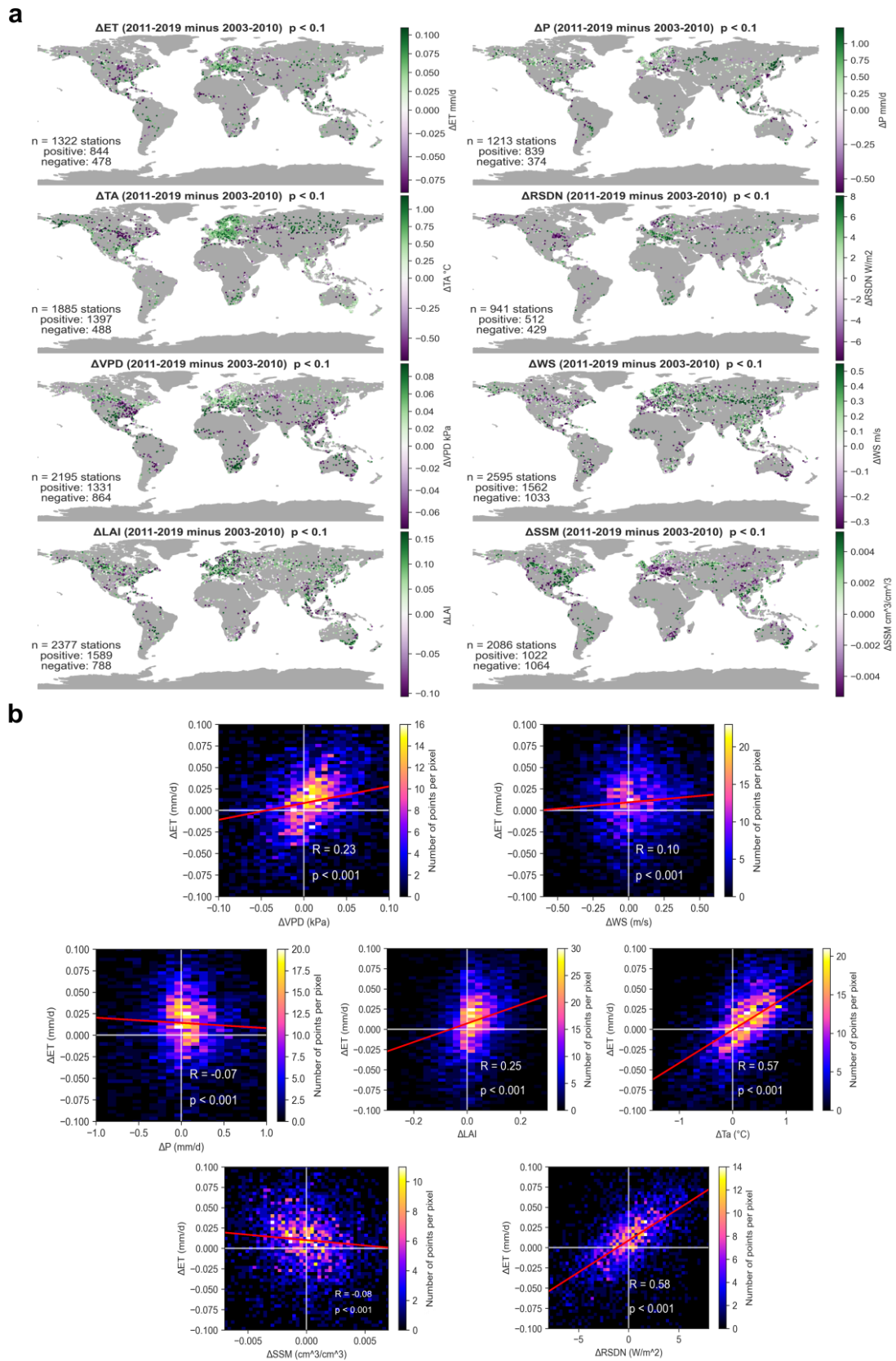

**Fig. 4.** Significant (p < 0.1) changes in ET and environmental drivers from 2003-2010 to 2011-2019 at weather stations. **a**, spatial patterns of the significant changes. **b**, correlations between ET change and environmental driver changes. Ta, temperature. P, precipitation. RSDN, downward shortwave radiation.

VPD, vapor pressure deficit. WS, wind speed. LAI, leaf area index. SSM, surface soil moisture.

In addition to the above analysis via simple correlation analysis, to better understand the causal mechanism, via the casual graph model Bayesian Network (BN) (Fig. 5a), we investigated the sensitivity of ET changes (ΔET) between the period 2003-2010 and 2011-2019 to changes in climatic and environmental drivers using sensitivity analyses of BN (Fig. 5b). With the explicit causal relationship representation, BN can better account the total causal effect strengths (i.e., including both the direct and indirect effects) when there is the covariance between the variables and difficulty to distinguish between correlation and causality[34]. Mutual information (MI) is used as the sensitivity index (see Methods) to measure the strength of ΔET response to changes in climatic and environmental drivers. Globally, ΔET is very sensitive to ΔTa and ΔRSDN, followed by ΔLAI and ΔVPD. In Europe, ΔET is more sensitive to ΔSSM than in other continents. Compared to the results of the above correlation analysis, the contribution of ΔVPD relative to ΔLAI appears to have risen, suggesting that ΔVPD, in addition to increasing atmospheric water demand, can indirectly affect ΔET by influencing ΔLAI. This effect appears to be higher in the humid regions than in the drylands.

We also attributed significant ($p < 0.1$) changes in ET by analyzing changes in the probability of the drivers' changes in the context of ET's significant increases (Fig. 5c). In terms of the significant increase of ET (Fig. 5c), the contribution of the increase of Ta, RSDN, LAI, and VPD is considerable. The Ta increase is the common largest contributor to the ET increase. RSDN increase plays a stronger role in Europe. In the globe, the contributions of VPD and LAI increase are equivalent. This suggests that the effects of rising atmospheric water demand and greening are comparable (which incorporated the indirect effect of VPD on ET by controlling LAI). The contribution of the VPD increase is greater than the LAI increase in Asia and Europe, while the contribution of the LAI increase is greater than the VPD increase in North America. The contributions of VPD increase and LAI increase are also equivalent in drylands, while the contribution of VPD increase is slightly higher than LAI increase in humid regions. Compared to drylands, the contribution of WS increase is higher in the humid regions. The SSM decrease also contributed to the ET increase in humid regions. It is probably due to the higher atmospheric water demand that has led to both the ET increase and SSM decrease, and the LAI increase also allocated more surface water to vegetation transpiration rather than to SSM. In contrast, in drylands, the weak increase in SSM contributed to ET increase, probably due to the increased irrigation at some dryland stations leading to a synchronized rise in SSM and ET. The relatively small contribution of P change to ET increase can be caused by the partial absence of P observations at the stations in this study, thus limiting the contribution of P on ET changes. Fortunately, the inclusion of SSM changes may have compensated for the effects of changes in surface water supply which P often represented. Besides, other factors may also contribute directly or indirectly by affecting the above drivers first such as land cover change[15], changes in groundwater levels[39], fertilization effects of increasing $CO_2$ concentration[40], nitrogen deposition[15], and changes in agricultural practices[35] relevant to human activities. The limitations of spatial and temporal information of these factors have constrained the accurate quantification of these impacts which can be partially reflected by effects of variables such as LAI.

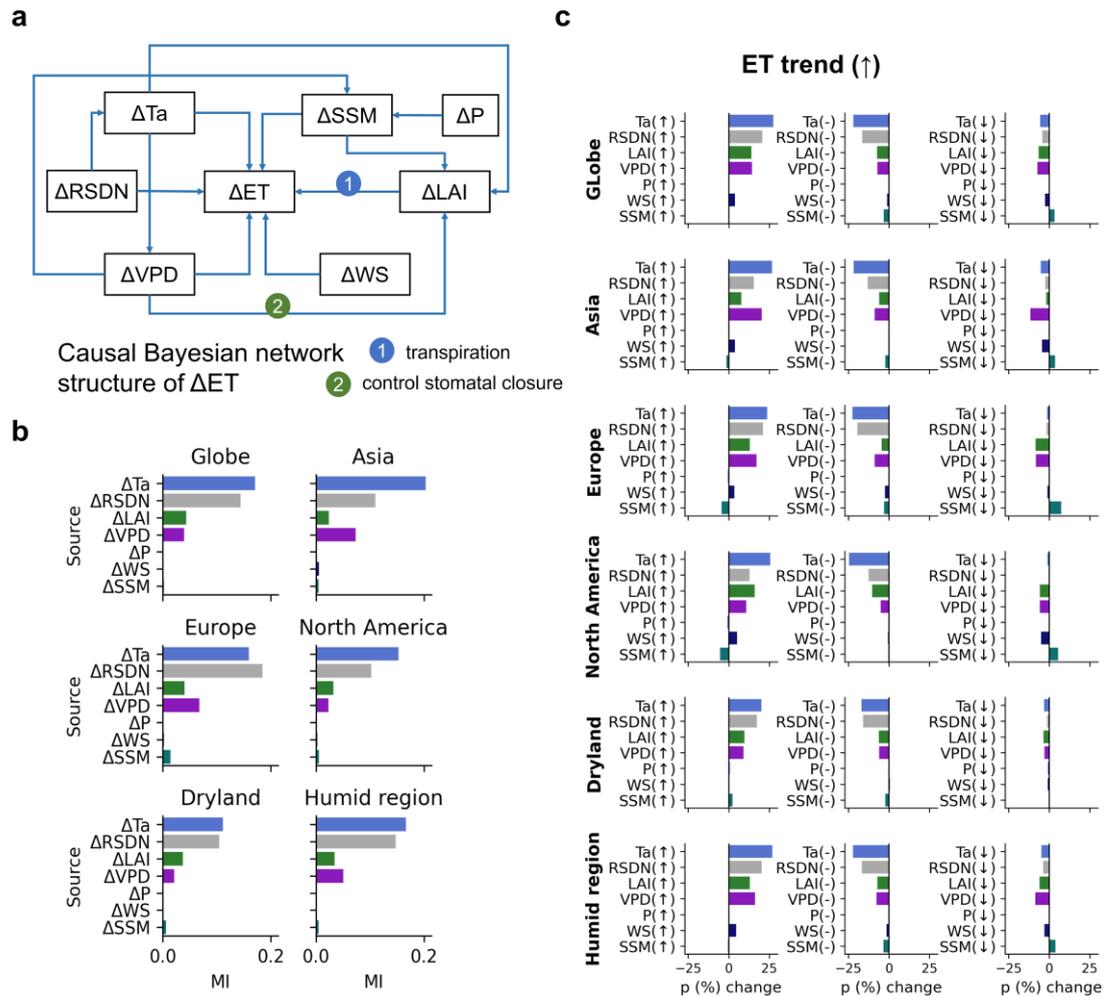

**Fig. 5.** Attribution of ET changes (between the two periods 2003-2010 and 2011-2019) by the Bayesian Network (BN). **a,** The Bayesian network structure for analyzing the causality of ΔET (between the two periods 2003-2010 and 2011-2019) and changes in environmental drivers. **b,** Sensitivity of ΔET to ΔVPD, ΔLSWI, ΔLAI, ΔWS, ΔTa, ΔP, and ΔRSDN based on the constructed BN. MI, mutual information (see Methods). Ta, air temperature. P, precipitation. RSDN, downward shortwave radiation. VPD, vapor pressure deficit. LAI, leaf area index. SSM, surface soil moisture. WS, wind speed. **c,** The change in the probability p (%) of 'significant increase' (↑), 'insignificant change' (-), and ' significant decrease' (↓) of the drivers in the context of a significant increase in ET compared to the baseline.

## Discussion

Although various methods have been used for global ET estimation, their dependence on empirical parameterization, the lack of global accuracy, and low forcing climate data resolution limit our understanding of global ET patterns and dynamics. The findings of this study were derived from data-driven station-scale records, thus reducing the negative impact of the empirical representation of processes, especially water stress, which is empirically represented in all other products. This empirical representation leads to inherent uncertainties of changes in the water cycle.

From 2003 to 2019, we obtained a global ET value range (493 to 522 mm/yr) and an increasing trend in global ET (0.60 mm yr$^{-2}$). This is comparable to the values given in previous studies. When using station averages to represent large regions, ET_WS also exhibits comparable reliability at the basin scale to gridded ET data such as ET_FLUXCOM and ET_MOD16. The reliability of ET_WS is even higher in basins with relatively high station density and representativeness. It is promising for ET-related studies in these regions with dense weather stations. When SSM, LAI, and other variables are used together, we can obtain a broader and more accurate picture of the complex response of ET to climate change. Further, a recent study[41] has suggested the feasibility of estimating carbon fluxes at weather stations. Combining ET and carbon flux data from global weather stations can greatly improve our understanding of issues such as land-atmosphere interactions, ecosystem water use efficiency, and carbon-water coupling mechanisms. These station-scale findings will also play an important role in guiding the management of local ecological conservation and agricultural activities. We will be able to regulate the characteristics of ecosystem response to climate change in a wide range of regions based on weather stations.

In terms of the ET change mechanism, from a decadal perspective, Ta, RSDN, VPD and LAI increases are the most significant contributors to ET increase. The positive effects of elevated Ta, RSDN, and VPD on ET were widespread and together exhibited higher atmospheric water demand. The mechanisms of LAI increase inducing ET changes can be diverse, which include extension of the growing season due to increased Ta, changes in vegetation cover and stomatal conductance due to increased $CO_2$ concentration, changes in the plant hydraulic characteristics due to VPD and soil moisture change, nitrogen deposition, and changes in land cover, etc. All of them may indirectly contribute to ET through LAI, and the mechanism of ET change due to increased LAI may be different in areas with different vegetation cover. In high vegetation areas, the positive effect of $CO_2$ fertilization on ET may be partially offset by the decrease in stomatal conductance caused by the increase in $CO_2$ concentration. In contrast, the positive effect of $CO_2$ fertilization on ET may be dominant in low vegetation areas. It demonstrated the different impacts of elevated $CO_2$ concentration. At seasonal scales, these mechanisms also showed heterogeneous patterns, with relatively greater changes likely to occur in spring and autumn. In addition, BN-based causal analyses indicated that the contributions of changes in VPD and LAI to ET changes were comparable globally. The BN included four pathways by which VPD affects ET: (i) the direct effect of ΔVPD on ΔET, (ii) the indirect effect of ΔVPD on ΔET through ΔLAI, (iii) the indirect effect of ΔVPD on ΔET through ΔSSM, (iv) the indirect effect of ΔVPD on ΔET through ΔSSM and then through ΔLAI. Therefore, the importance of ΔVPD is considerable when the major direct and indirect effects of ΔVPD on ΔET are included. It suggests that a comprehensive incorporation of the various causal pathways is necessary when quantifying the effects of climate change variables on ET[34]. The decadal-scale effects of water supply represented by SSM are relatively small, but this does not negate the potentially important role of SSM on an interannual scale[5].

Given that ET_WS and the global grids' average are generally comparable at the basin scale, the attribution of ET changes based on weather station ET and driver data can also be seen as an additional perspective to previous studies on attribution analysis[5,6,15,25]. Typically, weather stations are densely distributed in areas with dense human footprints, thus providing a good representation of ET in these regions. However, they are relatively sparsely distributed in regions such as deserts and tropical rainforests in Africa and South America. Given that these regions play an important role in the global water and carbon cycle, it is necessary to increase the density of weather stations in the future. In these regions, dense placement of flux stations is not feasible due to the extremely high costs. However, the cost of installing new weather stations is much lower and it will greatly improve our knowledge of the ET-ecosystem-climate relation in these regions when remote sensing data and machine learning are further combined to obtain station-scale ET data.

In summary, it is important to enhance the accuracy of our global ET estimates, and our study shows great potential for using ET estimates inferred from weather stations. Previous analyses on the water cycle and ET-ecosystem-climate relations based on coarse-resolution gridded ET data driven by coarse climate reanalysis can be revisited by using ET from weather stations as this study proposed. These future studies are likely to update much of the previous understanding related to ET, ET-ecosystem-climate relation, and the terrestrial water cycle.

## Methods

**ET predictions at weather stations**

We applied the ET estimation models developed from the flux station data to the weather stations in this study. Meteorological variables (in the FLUXNET2015 dataset), and biophysical variables extracted from MODIS imagery are used to predict daily ET with Random forests (RF) because it was found to outperform other algorithms[42]. The daily latent heat (LE) aggregated from half-hourly observations of the FLUXNET2015 dataset was used as the predicted variable. To control for potential biases and errors caused by gap-filling the FLUXNET data processing, LE observations with a quality control flag of LE (i.e. LE_F_MDS_QC in the FLUXNET2015 dataset) lower than 0.75 were excluded[43]. A total of 7 predictors controlling ET at a daily scale were used in the prediction model: daily mean temperature (Ta), daily temperature range[26] (TArange) calculated as maximum half-hourly temperature (TAmax) minus minimum half-hourly temperature (TAmin), wind speed (WS), vapor pressure deficit (VPD), downward shortwave radiation (RSDN), surface soil moisture (SSM) and LAI. We extracted LAI and RSDN at a scale of 500 m for both flux and weather stations (from the Global Daily Summary weather station observation dataset (GSOD)) after quality control from various data sources (Supplementary Note. 1). VPD is calculated from TAmax, TAmin, and dew point temperature (Tdew)[44]:

$$VPD = \frac{0.6108}{2}\left[exp\left(\frac{17.27 TAmax}{TAmax + 237.3}\right) + exp\left(\frac{17.27 TAmin}{TAmin + 237.3}\right) - 2exp\left(\frac{17.27 Tdew}{Tdew + 237.3}\right)\right] \quad (1)$$

The RF model was constructed and evaluated for accuracy by a 'leave-one-site-out' cross-validation[4,45]. Most of the cross-validations showed high accuracy. TA, RSDN and LAI showed higher feature importance in the used predictors (Supplementary Fig. 7). Subsequently, daily ET records were successfully estimated for over 8000 GSOD weather stations. To assess the predictability of weather stations (applicability of the RF model) and screen for more accurate ET estimates, we projected the R level (i.e., 'R > 0.7' and 'R <= 0.7') for each weather station. The R projection step was based on the relationship between the corresponding R records and distance $d$ defined[31,32] as the mean of the differences between the value of each predictor in the test set and the nearest 10 values of the predictor in the training set (Supplementary Note. 2) of all predictors in the cross-validation. This R projection model is trained by an RF classifier. It has a total classification accuracy of 79% (Supplementary Fig. 8) and thus can provide a relatively reliable R-value level projection for each weather station. Consequently, a total of 6277 weather stations with 'R > 0.7' (Supplementary Fig. 8) were used in the ET trend and pattern analysis.

**The monthly and annual mean of ET**

To calculate the interannual mean ET, for each year in 2003-2019 of each station (the year 2002 was excluded), data with more than 270 days of available ET records in a year were screened out (i.e., daily-scale ET at a station for a given year with records less than 270 were considered as ineffective representative of the year's average value). In the BN-based attribution of ET changes, the annual P, Ta, RSDN, LAI, VPD, SSM, and WS are also constrained by this 270-day criterion. For monthly ET, data with more than 22 days of available ET records in a month were considered ineffective representatives of the month's average value and screened out.

Due to the difference in the degree of representation of the actual area of various land cover or climate types by global weather stations, the direct averaging of ET trends from weather stations to calculate the annual global ET values may have a large bias. Therefore, in the calculation of the global average ET trend, we weighted the estimated ET of weather stations by the weights of the actual area of each land cover type to obtain the global mean ET with a reduced bias. It was named ET_WS and further compared with the trends derived from other global ET products. In this step, the plant functional types (PFTs) of the weather stations were derived from the 500 m resolution land cover data MCD12Q1 from MODIS. The most frequent type of multi-year land cover types (containing PFT information) for each site was extracted during its monitoring period.

**Other ET products for the comparison**

We compared the interannual dynamics reported by four ET products based on land surface models or remote sensing. We calculated the annual z-score for each product and also the ET predicted in this study. The ET products used for comparison include (1) ET from GLDAS2.2[46] which includes Data Assimilation from GRACE at a daily scale, and 0.25°. (2) ET from MOD16A2GF version 6 (ET_MOD16) using a Penman-Montieth-based method[16]. It is available from 2000 to the present at the 8-day temporal resolution and spatial resolution of 500 m. (3) ET from a product calculated with the PT-JPL algorithm with 36-km resolution[47]. (4) ET from FLUXCOM[4] (ET_FLUXCOM) including the remote-sensing-based product (RS) and the RS and meteorologically derived product (RS + METEO) are averaged into one estimate.

**Validation of ET via basin water balance**

For the validation of the estimated ET, we used the monthly P, Q, and TWS of large river basins based on the water balance budget. Basin ET (ET_WB) should typically equal P minus the terrestrial water storage change (ΔTWS) and Q: ET = P - Q - ΔTWS. ET_WB can be regarded as independent ET data at the basin scale. To evaluate the ability of ET_WS to represent ET at regional scales relative to ET_FLUXCOM and ET_MOD16, Pearson's correlation coefficient between ET_WS and ET_WB was compared with the correlation coefficient between ET_WB and ET_FLUXCOM and the correlation coefficient between ET_WB and ET_MOD16. The basin-scale P data was from the Global Precipitation Climatology Centre (GPCC)[48] based on P gauges. Q data was from a runoff dataset Global Runoff (GRUN)[49] based on streamflow gauges and machine learning, and TWS anomalies data was from the Gravity Recovery and Climate Experiment (GRACE)[50]. In global 189 basins[51], 53 basins used were selected in the validation by the following two conditions: (i) the weather station density (station number divided by basin area) should be over the 25th quartile (i.e., 33 stations per million square kilometers), (ii) the sum of the area of land cover types represented by the weather stations should be over 70% of the total basin area.

**Attributing ET variations and changes with BN**

A BN[52] is a directed acyclic graph used to model causality. It was found effective in representing causality beyond correlations and reducing the interferences from covariance between variables when used to compare predictor importance in ecosystem function driving-mechanism analysis[34]. It comprises nodes representing discrete or continuous quantities and directed edges which do not form a directed cycle. The nodes of a BN represent random variables ($X_1,., X_n$) with their joint probability distribution calculated as:

$$P(X) = P(X_1, X_2, \dots, X_n) = \prod_{i=1}^{n} P(X_i | pa(X_i)) \qquad (2)$$

where pa ($X_i$) represents the value of the parent node of the node $X_i$.

Here we first constructed a causal structure (Fig. 5a) of ΔET between the two periods 2003-2010 and 2011-2019 with ΔP, ΔTa, ΔRSDN, ΔVPD, ΔWS, ΔSSM, and ΔLAI to represent their relations. The BN incorporates both meteorological factors and underlying conditions (i.e., LAI and SSM). It can highlight the possible contribution of surface vegetation and moisture information extracted from remote sensing to changes in ET, including the indirect effect of meteorological variables affecting surface vegetation conditions and thus ET indirectly[34]. Each node contains three levels (i.e., 'significant positive change', 'insignificant change', and 'significant negative change' between the two periods 2003-2010 and 2011-2019).

Then, using the expectation-maximization algorithm[53], we compiled this BN and updated the conditional probability table with global and continental-scale ΔET, ΔP, ΔTa, ΔRSDN, ΔVPD, ΔWS, ΔSSM, and ΔLAI at weather stations. Subsequently, we performed the sensitivity analysis on the ΔET node to other nodes to understand the possible driving mechanisms of ΔET and the regional differences in the context of global warming and climate change. Specifically, to evaluate the sensitivity of a target node to other nodes, sensitivity analysis[54,55] of the BN is often based on the mutual information (MI) representing the entropy reduction of the distribution of the child node caused by the status change of the parent node. The greater the MI value, the higher the sensitivity of the child node to the parent node, accompanied by stronger causality. The formula of MI is as follows:

$$\text{MI} = H(Q)-H(Q|F) = \sum_q \sum_f P(q, f) \log_2 \left(\frac{P(q,f)}{P(q)P(f)}\right) \qquad (3)$$

where H represents the entropy, Q represents the target node, F represents the set of other nodes and q and f represent the status of Q and F.

In addition, based on the diagnostic analysis function of BN, we also recorded the change in the probability of various statuses of the driver nodes when the ΔET node is determined as 'significantly increase'. In this way, we can evaluate the contribution of the specific increase and decrease of the driver nodes to the increase and decrease of ET by comparing the probability change.

**Acknowledgments**

This research was supported by the National Natural Science Foundation of China (Grant No. U1803243), the Key projects of the Natural Science Foundation of Xinjiang Autonomous Region (Grant No. 2022D01D01), the Strategic Priority Research Program of the Chinese Academy of Sciences (Grant No. XDA20060302), and High-End Foreign Experts Project. Gentine acknowledges funding from the National Science Foundation Center for Learning the Earth with Artificial Intelligence and Physics (LEAP), award # 2019625.

**Data availability**

The flux data are from FLUXNET2015 (https://fluxnet.org/data/fluxnet2015-dataset/), remote sensing data are derived from Google Earth Engine, and the global weather station data are from the weather records of the Global Daily Summary (GSOD), which are available from the National Centers for Environmental Information (NCEI) (www.ncei.noaa.gov/data/global-summary-of-the-day/archive/).

**Code availability**

All data and code support for this study can be available upon request from the first author (shihaiyang16@mails.ucas.ac.cn).